\documentclass{article}
\usepackage{spconf,amsmath,amssymb,graphicx,booktabs}
\usepackage[hypertexnames=false]{hyperref}
\usepackage{multirow}
\usepackage{float}
\usepackage{placeins}
\hypersetup{hidelinks}
\graphicspath{{figures/}}

\makeatletter
\def\thebibliography#1{%
  \section{References}%
  \list{[\arabic{enumi}]}{%
    \settowidth\labelwidth{[#1]}%
    \leftmargin\labelwidth
    \advance\leftmargin\labelsep
    \usecounter{enumi}}%
  \def\newblock{\hskip .11em plus .33em minus .07em}%
  \sloppy\clubpenalty4000\widowpenalty4000
  \sfcode`\.=1000\relax}

\makeatother

\title{SPARSE-LDV NEURAL RECONSTRUCTION WITH VIBRO-ACOUSTIC EVIDENCE FUSION FOR EARLY BOLT PRELOAD LOSS}
\name{Berkay Kullukcu, Dina Hannebauer \thanks{This work was funded by the European Regional Development Fund (ERDF/EFRE) and the State of Brandenburg within the StaF-Verbund programme, under the project ``Systementwicklung für intelligente und automatisierte mobile Inspektion von Schienenfahrzeugen (SIAMIS)'', administered by the Investment Bank of the State of Brandenburg (ILB).}}
\address{FG Machine Dynamics and Acoustics, TH Wildau, Hochschulring 1, Wildau, 15745, Brandenburg, Germany}

\begin{document}
\ninept
\maketitle

\begin{abstract}
Repeated dense laser Doppler vibrometry (LDV) is costly, while early bolt-preload loss may produce weak and resonance-dependent changes. We combine commissioning-baseline-assisted sparse-LDV reconstruction with force-normalized microphone evidence. At $m=24$, each single-bolt 5-Nm state is reconstructed by a neural-network fold that excludes that state from training. A healthy-calibrated late-fusion rule combines the hidden-node upper-tail local-FRAC deficit with microphone $H_1$ changes normalized by impact-level scatter. Sparse LDV alone flags 24/28 mild condition--resonance windows and acoustic evidence flags 25/28; fusion flags all 28/28. All four LDV misses occur at RG3 and are recovered acoustically. Fusion remains 28/28 for acoustic thresholds from 3 to 10 pooled standard deviations. The result supports complementary vibro-acoustic evidence for early preload-loss screening while avoiding pointwise fusion of separately acquired campaigns.
\end{abstract}


\section{Introduction}
Reduced bolt preload modifies contact stiffness, damping, and microslip and can precede gross joint failure, motivating methods that detect small changes before complete loosening \cite{huang2022,wang2019impact}. Scanning LDV provides spatially resolved complex frequency-response functions (FRFs), but repeated dense scans are expensive \cite{castellini2006}. Microphones add a non-contact channel sensitive to radiated structural response, yet a single sensing modality can have frequency-dependent blind spots \cite{becht2019}.

We retain one dense all-tight LDV scan as a commissioning baseline and reconstruct later fields from graph-selected sparse locations. The LDV database contains 18 torque configurations and seven previously screened resonance groups \cite{kullukcu_mssp,kullukcu_data}; a separate two-microphone campaign contains all-tight, four single-bolt 5-Nm states, and five severe states. This paper asks whether (i) an unseen 10-to-5-Nm single-bolt reduction remains detectable after sparse reconstruction and (ii) condition-level acoustic evidence can recover resonance windows in which sparse LDV is weak, without treating separately acquired LDV and acoustic experiments as synchronized pointwise pairs.

\section{Relation to Prior Work}
Sparse vibration-field recovery has progressed from graph-signal sampling and QR sensor selection \cite{shuman2013,anis2016,kim2021} to compressed-sensing and deep-learning reconstruction \cite{yuan2024sparse,dong2023dcgan,an2024sparsityfree}. Dense-to-sparse commissioning is particularly relevant: Zhang et al. use dense controlled measurements to support later sparse reconstruction \cite{zhang2025dictionary}. Our structural path follows this operational idea but retains the measured healthy field directly in the local-FRAC diagnostic reference.

Multimodal work establishes a complementary role for sound. Huang et al. fuse ultrasonic and audio information for threaded-fastener loosening \cite{huang2023multimodal}, while Wang et al. report that vibration-only sensing misses a substantial fraction of initial loosening and that vibration--microphone--AE fusion improves early-stage detection \cite{wang2025fusion}. More generally, decision-level multimodal fusion can outperform a single modality for progressive damage \cite{yang2024decision}, and evidence fusion has been used specifically to improve slight-damage sensitivity under noise \cite{guo2018multiscale}. These studies motivate our change from resonance reweighting to late complementary evidence fusion: LDV and microphones produce separate condition--resonance evidence scores, which are combined only at the decision level.

\begin{figure}[H]
\centering
\includegraphics[width=0.86\columnwidth]{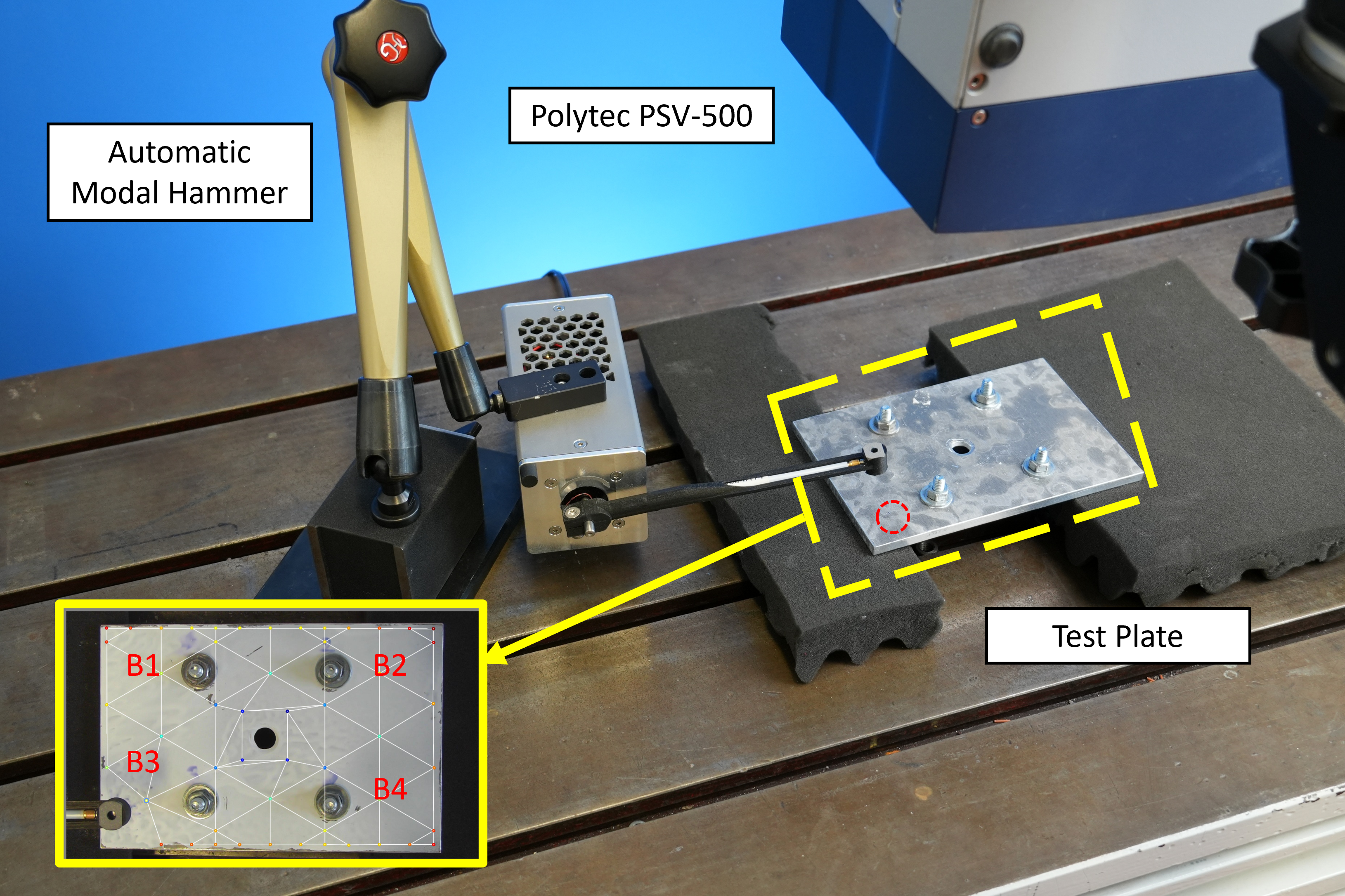}
\caption{Dense commissioning LDV campaign. The four-bolt plate is excited near B3 by the scalable automatic modal hammer (SAM) and scanned by the Polytec PSV-500 at 51 spatial locations.}
\label{fig:setup}
\end{figure}

\section{Method}
\subsection{Experimental Setups}
Figure~\ref{fig:setup} shows the scanning-LDV experiment. The $100\times150\times7$~mm aluminum plate is joined by four M6 bolts, supported on soft sponges, and excited near B3 by an automatic modal hammer. Complex $H_1$ displacement-to-force FRFs are available at 51 common locations. The seven retained resonance groups are centered at 3325, 4703, 7034, 7861, 8163, 8428, and 8695~Hz; each is represented by a tracked 121-bin ($\pm60$~Hz) window \cite{kullukcu_mssp}.

\begin{figure}[H]
\centering
\includegraphics[width=0.45\columnwidth]{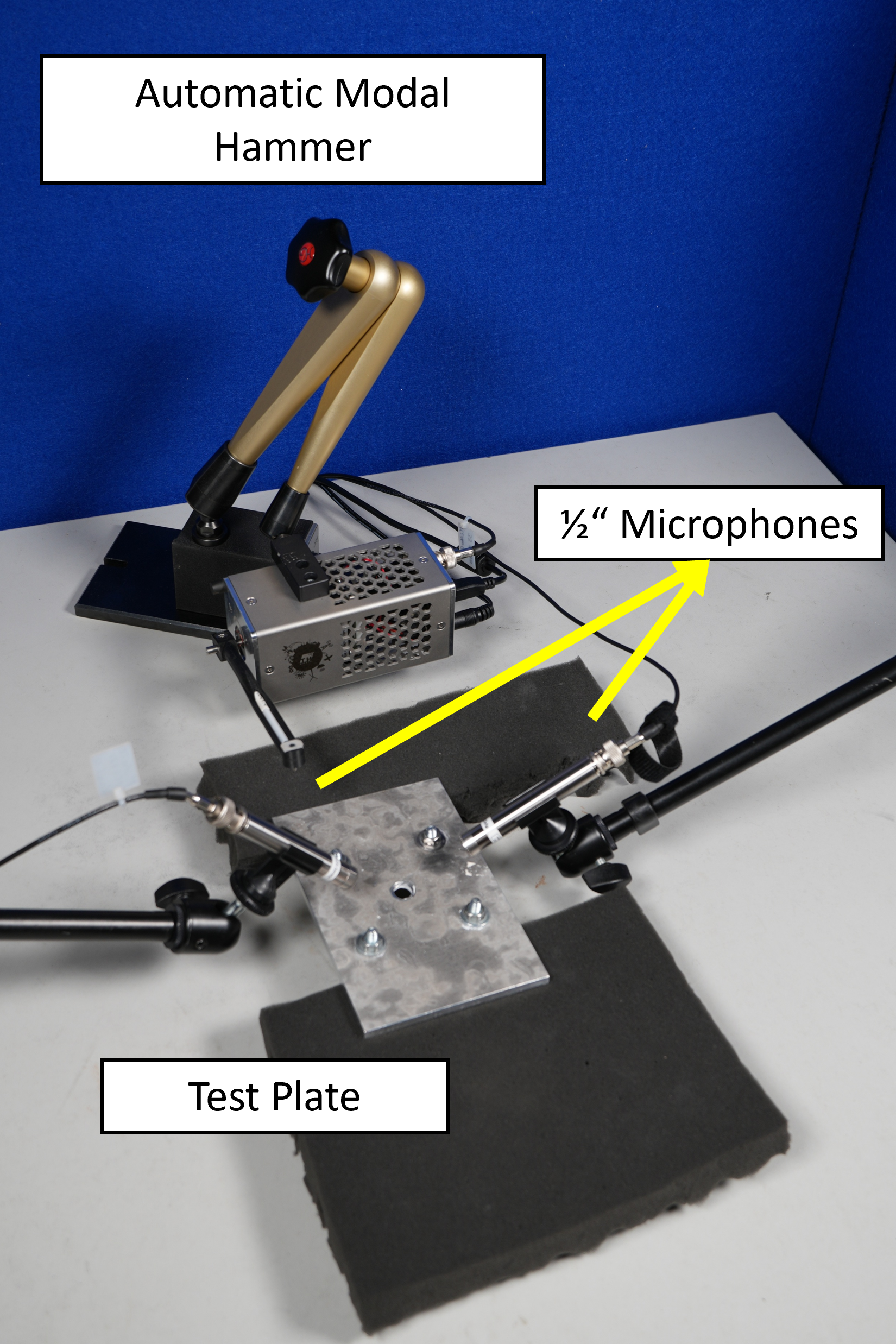}
\caption{Separate acoustic campaign used to estimate resonance reliability. Two matched $\frac{1}{2}$-inch free-field microphones are placed 30~mm from the B1--B2 and B3--B4 bolt pairs.}
\label{fig:acousticsetup}
\end{figure}

The separate acoustic campaign, Fig.~\ref{fig:acousticsetup}, uses two PCB 378B02 free-field microphones 30~mm from the B1--B2 and B3--B4 bolt pairs. The stereo microphone and hammer-force WAV files are sample-synchronous at 48~kHz. Ten states were measured: all-tight, four single-bolt 5-Nm states, all-loose, and four one-tight/three-loose states. After impact screening, the calibrated pressure-to-force estimator is
\begin{equation}
\widehat H_{pF,q}^{(1)}(f)=
\frac{\sum_{k=1}^{K}P_{q,k}(f)F_k^*(f)}
{\sum_{k=1}^{K}|F_k(f)|^2},
\label{eq:acoustic_h1}
\end{equation}
with microphone and hammer sensitivities applied to obtain Pa/N.

\subsection{Sparse Reconstruction and Diagnostic Maps}
The projected 51-point coordinates form a four-nearest-neighbor graph. Pivoted QR decomposition of the first $m$ normalized graph-Laplacian eigenvectors defines a fixed measurement set $\mathcal S_m$ \cite{anis2016,kim2021}. For each frequency bin, sparse observations initialize the field through
\begin{equation}
\widetilde{\mathbf h}=\arg\min_{\mathbf x}
\|\mathbf P_{\mathcal S}\mathbf x-\mathbf y\|_2^2+
\lambda\mathbf x^{\mathrm H}\mathbf L\mathbf x,
\label{eq:graph}
\end{equation}
followed by a residual MLP. Hard consistency restores measured rows exactly. Four fold-specific models are used: each 5-Nm state is evaluated by the fold in which that state was excluded from training. Complete- and multi-bolt-loose states are excluded from all fitting.

Using the dense measured all-tight baseline $H_b$, the local FRAC deficit at node $i$ is
\begin{equation}
D_i=1-\frac{|\sum_f H_{b,i}^*(f)H_{c,i}(f)|^2}
{\sum_f|H_{b,i}(f)|^2\sum_f|H_{c,i}(f)|^2}.
\label{eq:frac}
\end{equation}
Map fidelity is evaluated with Pearson correlation $r$, including a hidden-node metric that excludes the 24 exactly measured locations.

\subsection{Healthy-Calibrated Vibro-Acoustic Evidence Fusion}
\begin{figure*}[t]
\centering
\includegraphics[width=0.84\textwidth]{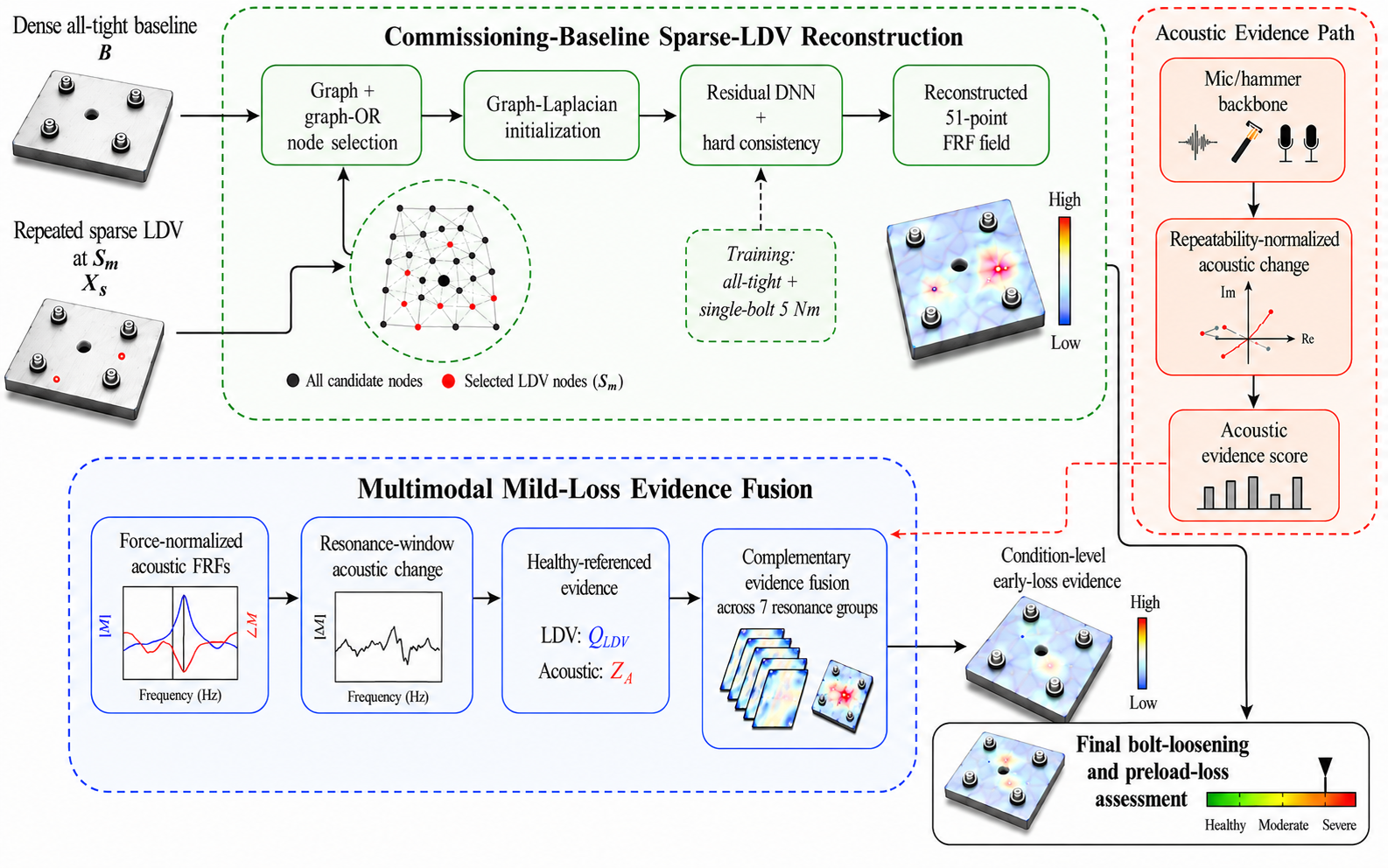}
\caption{Overview of the proposed multimodal workflow for bolt-loosening assessment. A dense all-tight commissioning baseline B is measured once and used to define the graph-based sparse-LDV reconstruction pipeline. Repeated sparse LDV measurements at the selected node set $S_m$ are expanded to a reconstructed 51-point FRF field using graph-based initialization, a residual DNN, and hard consistency. In parallel, hammer-normalized two-microphone measurements provide acoustic change features. The LDV-derived diagnostic information and the acoustic evidence are then combined to produce a condition-level diagnostic map and the final bolt-loosening assessment.}
\label{fig:workflow}
\end{figure*}

Figure~\ref{fig:workflow} summarizes the new early-loss analysis. Because loosening changes are spatially localized, structural evidence for condition $c$ and resonance $r$ is the 90th percentile of the reconstructed local-FRAC deficit over hidden nodes $\mathcal H$,
\begin{equation}
Q^{\rm LDV}_{c,r}=P_{0.90}\{D_{i,c,r}:i\in\mathcal H\},\qquad
\tau_L=\max_r Q^{\rm LDV}_{\mathrm{healthy},r}.
\label{eq:ldvevidence}
\end{equation}
Thus $E_L=Q^{\rm LDV}_{c,r}/\tau_L$ is calibrated only from the reconstructed healthy commissioning state.

For microphone $q$, let $\Delta L_{q,c,r}$ be the baseline-relative RMS change of the force-normalized $H_1$ response in the same $\pm60$-Hz band. We normalize this change by impact-level repeatability,
\begin{equation}
Z_{q,c,r}=\frac{|\Delta L_{q,c,r}|}
{\sqrt{(\sigma_{q,c,r}^2+\sigma_{q,b,r}^2)/2}},\qquad
Z^A_{c,r}=\tfrac12(Z_{12,c,r}+Z_{34,c,r}).
\label{eq:acevidence}
\end{equation}
A deliberately conservative primary acoustic threshold $\tau_A=5$ pooled-SD units is fixed without optimizing on the mild states; $\tau_A=3$ and 10 are sensitivity checks. Decision-level evidence fusion follows the complementary late-fusion principle \cite{yang2024decision,guo2018multiscale},
\begin{equation}
E_F=\max\!\left(\frac{Q^{\rm LDV}_{c,r}}{\tau_L},\frac{Z^A_{c,r}}{\tau_A}\right),
\qquad \text{flag if }E_F>1.
\label{eq:fusion}
\end{equation}
The max rule intentionally asks whether either independent modality provides convincing evidence; it does not use acoustic values to alter spatial LDV nodes. As a secondary condition-level analysis, a nested leave-one-condition-out ridge model estimates total torque loss from the median LDV upper-tail score, alone or together with the median absolute acoustic change.

\section{Results}
At $m=24$, repeated LDV acquisition is reduced from 51 to 24 locations (52.9\%). Across the 28 strictly out-of-fold mild condition--resonance maps (four single-bolt 5-Nm states $\times$ seven resonances), median full-map and hidden-node correlations are $r=0.940$ and 0.870. For comparison, the 13 unseen complete-/multi-loose states give 91 condition--resonance samples; Table~\ref{tab:budget} shows the severe-state budget dependence.

\begin{table}[H]
\caption{Dense-reference residual-DNN performance across 91 unseen severe condition--resonance samples; entries are medians.}
\label{tab:budget}
\centering
\setlength{\tabcolsep}{1.7pt}
\begin{tabular}{rcccc}
\toprule
$m$ & NMSE & Phase (deg.) & $r_{\rm full}$ & $r_{\rm hid}$\\
\midrule
4  & 0.263 & 10.86 & 0.619 & 0.613\\
8  & 0.092 &  6.61 & 0.648 & 0.616\\
12 & 0.120 &  6.89 & 0.703 & 0.583\\
16 & 0.065 & 5.90 & 0.820 & 0.708\\
24 & \textbf{0.035} & \textbf{5.05} & \textbf{0.871} & \textbf{0.731}\\
\bottomrule
\end{tabular}
\end{table}

\begin{figure}[H]
\centering
\includegraphics[width=0.80\columnwidth]{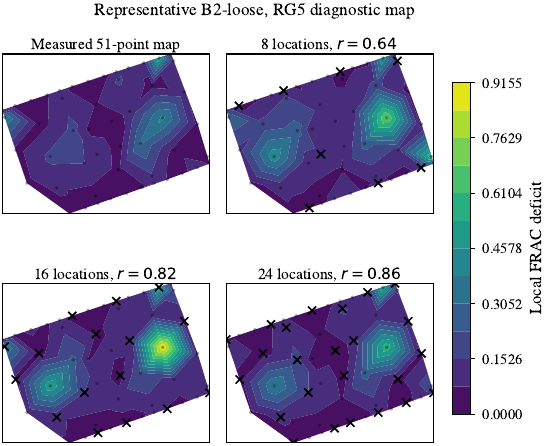}
\caption{Representative B2-loose RG5 local-FRAC deficit field. The dense measured map is compared with reconstructions from progressively larger sparse sets; all use the same measured 51-point healthy commissioning reference.}
\label{fig:maps}
\end{figure}

The acoustic input is highly repeatable: 58 impacts pass screening, median target-band coherence is 0.9947, and raw-WAV $H_1$ FRFs reproduce exported transfer functions with median complex FRAC 0.9991. A leave-one-impact-out check within the healthy state gives a maximum absolute band-RMS change of only 0.983~dB, supporting the use of baseline-relative acoustic evidence.

\begin{figure}[t]
\centering
\includegraphics[width=0.98\columnwidth]{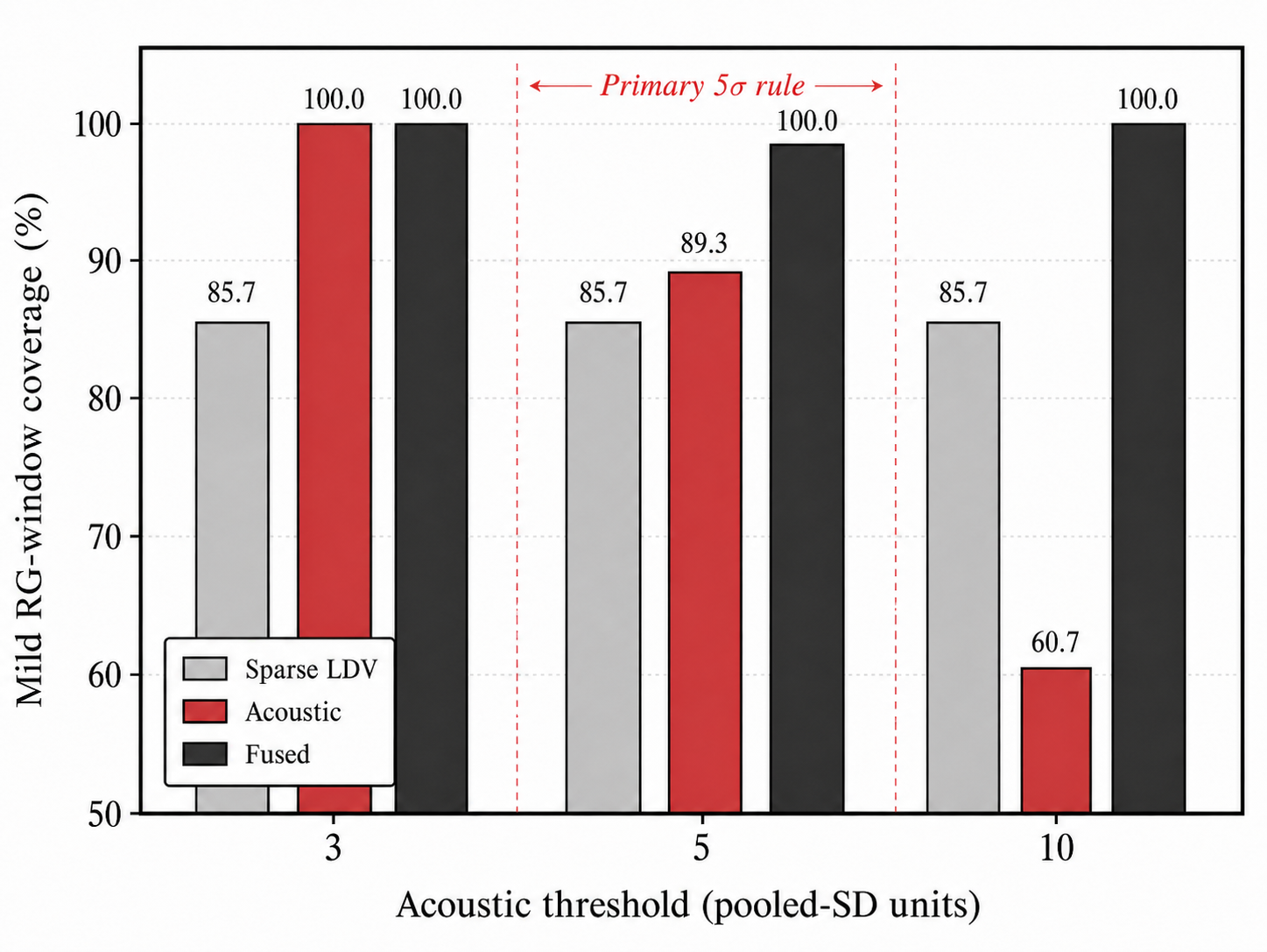}
\caption{Mild-condition resonance-window coverage for sparse LDV, acoustic evidence, and multimodal fusion as a function of the acoustic threshold. Results are shown for thresholds of 3, 5, and 10 pooled-standard-deviation units. The highlighted $5\sigma$ rule is used as the primary operating point. Sparse LDV alone covers 85.7$\%$ of the mild resonance-group windows, whereas acoustic-only coverage depends strongly on threshold choice. In contrast, the fused method maintains 100$\%$ coverage across all tested thresholds, showing that acoustic evidence recovers the mild-condition windows missed by sparse LDV.}
\label{fig:mildfusion}
\end{figure}

For early-loss screening, the healthy LDV floor is $\tau_L=0.0517$. Sparse LDV alone exceeds this floor in 24/28 mild resonance windows (85.7\%). With the primary $5\sigma$ acoustic rule, microphones independently flag 25/28 (89.3\%). The decision-level fusion flags 28/28 (100\%) [Fig.~\ref{fig:mildfusion}(a)]. All four LDV misses occur at RG3, one for each 5-Nm bolt condition; their acoustic standardized evidence is 14.4--26.4, so all are recovered by the microphone channel. Conversely, the one acoustic miss at the primary threshold is recovered by LDV. At the physical-condition level, LDV alone has all seven resonances above threshold in 0/4 mild states and acoustic evidence in 1/4, whereas fusion achieves 7/7-window coverage in all 4/4 states.

The result is insensitive to the chosen acoustic evidence threshold. Acoustic-only coverage decreases from 28/28 at $3\sigma$ to 25/28 at $5\sigma$ and 17/28 at $10\sigma$, while fused coverage remains 28/28 throughout [Fig.~\ref{fig:mildfusion}(b)]. These are resonance-window coverage values, not 28 independent specimen trials. As a secondary severity result, nested leave-one-condition-out ridge regression reduces mild-state torque-loss MAE from 1.15~Nm with LDV alone to 0.62~Nm with LDV+acoustic features (46.4\% reduction; three of four mild conditions improve); this small-$n$ result is exploratory.

\begin{figure}[t]
\centering
\includegraphics[width=0.98\columnwidth]{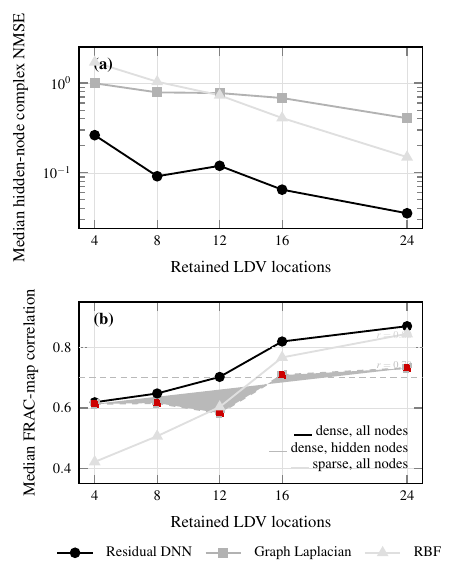}
\caption{Effect of retained LDV budget $m$. (a) Median hidden-node complex NMSE for the residual DNN and graph-Laplacian/RBF baselines. (b) Median FRAC-map correlation for dense-reference full and hidden maps and the sparse-reference protocol.}
\label{fig:budget}
\end{figure}

Figure~\ref{fig:budget} confirms that $m=24$ supplies the strongest structural basis for the early-loss fusion while still removing more than half of repeated scan locations.

\section{Discussion}
The principal new result is not that microphones replace LDV, but that the two modalities fail differently. Sparse LDV preserves the mild 5-Nm spatial field well overall ($r_{\rm hid}=0.870$ median), yet all four healthy-floor misses cluster at RG3. Force-normalized acoustic evidence is very strong precisely in those four windows. The max-evidence rule therefore raises cross-resonance mild coverage from 85.7\% to 100\% without training a fusion model on the mild states. The persistence of 28/28 coverage when the acoustic threshold is tightened from 3 to 10 pooled-SD units argues that the result is not tied to a single permissive threshold.

This behavior is consistent with the multimodal literature. Wang et al. report 63\% detection of initial loosening from vibration alone versus 94\% after adding microphone and AE information \cite{wang2025fusion}; Huang et al. similarly exploit complementary ultrasonic and audio features \cite{huang2023multimodal}. Our analysis differs in two ways. First, it retains a spatial LDV diagnostic field rather than replacing it with a classifier. Second, because the acoustic and LDV campaigns are separate realizations, we fuse only normalized evidence at the condition--resonance decision level. This is closer in spirit to late/confidence fusion \cite{yang2024decision} and evidence aggregation for slight damage \cite{guo2018multiscale} than to pointwise multimodal regression.

The use of a hidden-node 90th percentile is also deliberate: local bolt-related FRAC changes are hotspot-like, so a spatial mean can dilute them while a maximum is overly sensitive to one reconstructed node. Nevertheless, the present study has important limits. There is one specimen and one healthy acoustic assembly realization; the 28 windows are correlated within only four mild physical conditions; no multi-bolt 5-Nm acoustic states exist; and separate campaigns cannot quantify assembly-to-assembly false alarms. Therefore 100\% should be read as commissioning-set resonance coverage, not a population detection probability. Repeated reassemblies with simultaneous hammer, microphones, and sparse/dense LDV are required for calibrated sensitivity/specificity and mild multi-bolt validation.

\section{Conclusion}
A healthy-calibrated late-fusion strategy was added to commissioning-baseline-assisted sparse LDV for early bolt preload loss. With 24 of 51 LDV locations, four unseen single-bolt 5-Nm states retain high spatial map fidelity. At a conservative $5\sigma$ acoustic criterion, LDV-only and acoustic-only evidence cover 24/28 and 25/28 mild resonance windows, respectively, whereas decision-level fusion covers all 28/28; every LDV miss is recovered acoustically. Fused coverage remains 28/28 for acoustic thresholds from 3 to 10 pooled standard deviations. The data therefore support microphones as a complementary early-warning channel for resonance-dependent sparse-LDV blind spots, while larger repeated and synchronized experiments are needed for population-level detection claims.

\bibliographystyle{IEEEtran}
\bibliography{strings,refs,refs_priorwork}
\end{document}